\documentclass[epsf,twocolumn,showpacs,preprintnumbers]{revtex4-2}

\usepackage{amsmath}
\usepackage{graphicx}
\usepackage[colorlinks=true, allcolors=blue]{hyperref}
\usepackage{dcolumn}
\usepackage{bm}
\usepackage{ulem}  

\newcommand{\eups}{EuPd$_2$Si$_2$}
\newcommand{\eupg}{EuPd$_2$Ge$_2$}

\begin{document}\title{ Microscopic analysis of the valence transition in tetragonal 
\eups}

\author{Young-Joon Song$^{1}$}
\author{Susanne Schulz$^{2}$}
\author{Kristin Kliemt$^{3}$}
\author{Cornelius Krellner$^{3}$}
\author{Roser Valent\'{i}$^{1}$}

\affiliation{
 $^1$Institut f\"ur Theoretische Physik, Goethe-Universit\"at Frankfurt, 
 Max-von-Laue-Stra\ss e 1, 60438 Frankfurt am Main, Germany\\
$^2$Institut f\"ur Festk\"orper- und Materialphysik, 
Technische Universit\"at Dresden, 01062 Dresden, Germany\\
$^3$Kristall- und Materiallabor, Physikalisches Institut, 
Goethe-Universit\"at Frankfurt, Max-von-Laue Stra\ss e 1, 60438 Frankfurt am Main, Germany
}
\date{\today}

\begin{abstract}
Under temperature or pressure tuning, tetragonal \eups~is known to undergo
a valence transition from nearly divalent
to nearly trivalent Eu accompanied by a volume reduction.
Albeit intensive work, its microscopic origin is still being discussed.
Here, we investigate the mechanism of the valence transition under volume compression
by {\it ab initio} density functional theory (DFT) calculations.
Our analysis of the electronic and magnetic properties of \eups\ when approaching 
the valence transition shows an enhanced  $c$-$f$ hybridization 
between localized Eu 4$f$ states and itinerant conduction states 
(Eu 5$d$, Pd 4$d$, and Si 3$p$) where an electronic charge redistribution takes place.
We observe that the change in the electronic structure is intimately related to 
the volume reduction where Eu-Pd(Si) bond lengths shorten and, for the transition to happen, 
we trace the delicate balance between electronic bandwidth, crystal field splitting, 
Coulomb repulsion, Hund's coupling and spin-orbit coupling.
In a next step we compare and benchmark our DFT results to 
surface-sensitive photoemission data in which 
the mixed-valent properties of \eups{} are reflected 
in a simultaneous observation of divalent and trivalent signals 
from the Eu $4f$ shell. 
The study serves as well to explore the limits of density functional theory 
and the choice of exchange correlation functionals 
to describe such a phenomenon as a valence transition.
\end{abstract}
\maketitle

\section{Introduction}
For decades, 4$f$ electron systems have attracted much attention 
due to the realization of a large variety of interesting phenomena,
such as the Kondo effect and the emergence of heavy fermion features, quantum criticality, 
unconventional superconductivity, exotic magnetism, non-trivial topological phases, 
or valence transitions, to mention 
a few~\cite{wirth2016,rau2019,pfleiderer2009,stewart2001,onuki2020,onuki2017,xu2020,rahn2018}.
Valence transitions have been notably investigated in Eu-based systems where Eu can attain 
two possible valence states; divalent Eu$^{2+}$(4$f^{7}$) and trivalent Eu$^{3+}$(4$f^{6}$).
In divalent Eu$^{2+}$, following the Hund's rule in a $LS$ description,
seven electrons fill the $f$ states with a total orbital angular momentum $L$ = 0 
and a spin momentum $S$ = $\frac{7}{2}$, giving rise to a total angular momentum 
$J$ = $\frac{7}{2}$.
In trivalent Eu$^{3+}$, $L$ = $S$ = 3 and $J$ = 0.

Upon lowering the temperature a smooth change of the valency from a non-integer value 
close to a magnetic Eu$^{2+}$ state to a non-integer value 
close to a nonmagnetic Eu$^{3+}$ state has been observed, for instance, 
in tetragonal \eups~\cite{sampath}, EuCu$_{2}$Si$_{2}$~\cite{Bauminger,Patil}, 
and EuIr$_{2}$Si$_{2}$~\cite{chevalier,IrSi}. 
Alternatively, such valence transitions were also reported under the application 
of pressure in  tetragonal antiferromagnetic (AFM) EuRh$_2$Si$_2$~\cite{mitsuda3}, 
EuNi$_2$Ge$_2$~\cite{nakamura}, and EuCo$_2$Ge$_2$~\cite{dionicio}.
While the valence transition is expected to be related  
to structural and chemical bonding changes between localized 4$f$ 
and the more itinerant $s$, $p$ and $d$ electrons under temperature or pressure effects,
a full microscopic description of the transition mechanism is still lacking.

We focus here on the valence transition in \eups\ and provide a microscopic
analysis of the transition
by a combination of density functional theory (DFT) based calculations
and photoemission spectroscopy measurements.
We find that, when approaching the valence transition
upon volume compression, an enhanced 
$c$-$f$ hybridization between localized Eu 4$f$ states and 
itinerant conduction states (Eu 5$d$, Pd 4$d$, and Si 3$p$) happens 
where an electronic charge redistribution takes place.
We observe that the change in the electronic structure is intimately related to 
the volume reduction where the Eu-Pd(Si) bond lengths shorten and, 
for the transition to happen, 
we trace the delicate balance between electronic bandwidth, crystal field splitting, 
Coulomb repulsion, Hund's coupling and spin-orbit coupling.
The study serves as well to explore the limits of density functional theory 
and the choice of exchange correlation functionals to describe such a phenomenon 
as a valence transition.

The paper is organized as follows. 
In Sec. II we provide a summary of the known properties of \eups.
In Sec. III we describe the methods used for our study. 
In Sec. IV we present a comparative analysis of structural details 
in mixed-valence tetragonal Eu compounds Eu$TM_2X_2$ where $TM$ denotes 
the transition metal ion and $X$ = Si, Ge. 
In Sec. V we present our results on the electronic properties of \eups~for bulk 
and slab calculations and compare to experimental photoemission measurements.
Finally, in Sec. VI we present our conclusions.

\section{E\lowercase{u}P\lowercase{d}$_2$S\lowercase{i}$_2$} 

\begin{figure}[t]
\vskip 2mm
\includegraphics[width=1.0\columnwidth]{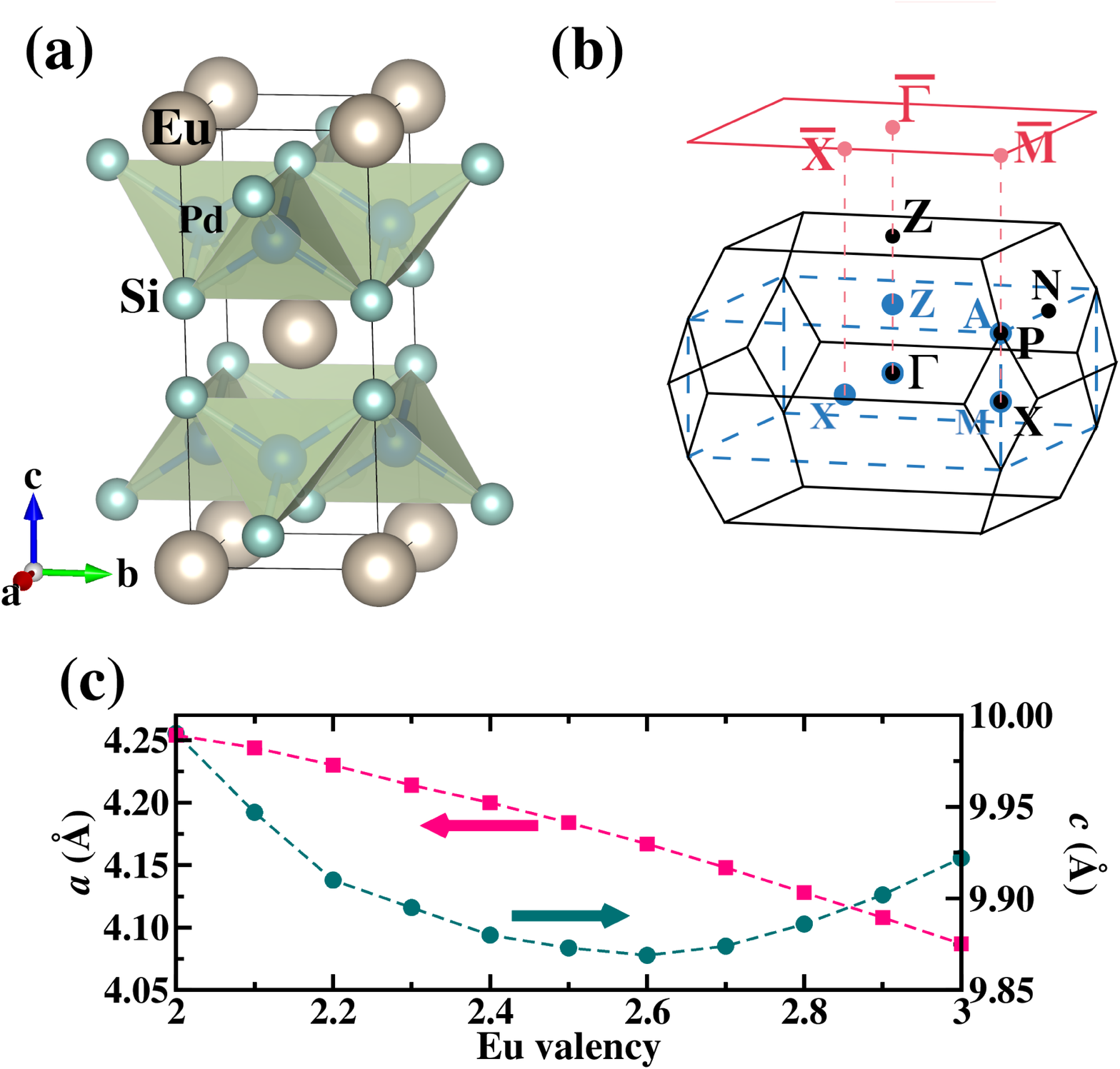}
\caption{ (a) Crystal structure of tetragonal \eups, produced by VESTA~\cite{vesta}. 
(b) First Brillouin zone of the tetragonal $I4/mmm$ (black, solid) 
and $P4/mmm$ (blue, dashed) structure with special $k$ points for the band structure. 
The (001) surface was projected in the slab calculations, 
which is drawn by a pink rectangle. 
(c) Fully relaxed lattice parameters of \eups~at each Eu valency
obtained using the {\it open-core} approximation within LDA.  
}
\label{str}
\end{figure}

The valence transition in mixed-valence tetragonal \eups~was first reported  
in 1981 in Ref.~\cite{sampath} where a strong isomer shift in M\"ossbauer spectroscopy
was observed when lowering the temperature from 200 K to 150 K
accompanied by an increase in the Eu average valency~\cite{sampath,mimura,mitsuda2}.
The Curie-Weiss moment of \eups~at high temperatures was reported 
to be 8.04{ }$\mu_{B}$~\cite{mitsuda1}, 
which is very close to the theoretically calculated value of 7.94{ }$\mu_{B}$
for the divalent state ($J$ = $\frac{7}{2}$).
It was then shown by means of x-ray absorption spectroscopy~\cite{nagarajan1981,wortmann1985}, 
M\"ossbauer spectroscopy\cite{wortmann1985}, and 
photoemission spectroscopy~\cite{mimura} measurements
that at a temperature of about 200{ }K \eups~enters into a valence crossover regime 
with the mean Eu valency changing smoothly from 2.3 to 2.8 between 200 and 100{ }K. 
Interestingly, within the valence crossover the lattice parameter 
$a$ = 4.24{ }\AA~at room temperature reduces below 50{ }K
to $a$ = 4.16{ }\AA, whereas $c$ remains 
unchanged~\cite{palenzona,mitsuda2,kliemt2022} [Fig.~\ref{str}(a)].

Further, applying high pressure can result in \eups~being 
in more mixed-valent states.~\cite{srinivasan1984,serdons2004,adams}. 
The authors of Ref.~\cite{adams} argued that the valence transition in \eups~would 
also occur under application of an external pressure of about 2{ }GPa, 
when $a$ reduces to about 4.16{ }\AA, which is similar to the value of $a$ 
in the low-temperature regime; specifically, Serdons $et$ $al.$ measured that the mean 4$f$ valence 
reached about 2.55 at 2{ }GPa and saturated to about 2.65 at around 5{ }GPa.~\cite{serdons2004}
Besides, isostructural and isovalent \eupg~which has a larger 
unit cell volume than \eups~doesn't undergo 
a valence transition at low temperatures but an AFM transition 
at $T_\mathrm{N} \approx 17${ }K~\cite{onuki2020}. 
Actually, the tunability of the Eu valence state by applying chemical 
and finite hydrostatic pressure was reported in 
single crystals of EuPd$_2$(Si$_{1-x}$Ge$_x$)$_2$~\cite{wolf2022}. 
At $x$ = 0.2 the long-range AFM order of Eu moments observed 
below $T_\mathrm{N} \approx$ 47{ }K is suppressed 
by the application of the hydrostatic pressure of 0.1{ }GPa inducing 
an intermediate valence state in Eu.
All these observations point to the important role of the structural changes 
under temperature or pressure affecting the Eu valency.

\section{Methods}
We performed DFT calculations using the full-potential all-electron codes WIEN2k~\cite{w2k}, 
and FPLO~\cite{fplo1,fplo2}. 
The first code considers a linear augmented plane wave basis to solve the Kohn-Sham equations, 
while the second is based on a local-orbital minimum basis. 
The exchange-correlation functional was treated within 
the local (spin) density approximation [L(S)DA] in both WIEN2k and FPLO codes. 
Crystal bulk structures of \eups~were fully relaxed in the tetragonal space group $I4/mmm$
within LDA using the {\it open-core} approximation as implemented in FPLO
until forces were smaller than 1 meV{ }\AA.
In this approximation the Eu 4$f$ states are removed from the valence basis 
and enter the bulk description as core orbitals, while fixing 
the mean 4$f$ occupancy $n$ to a given value. 
The value of $n$ was considered in steps of 0.1 from 6 to 7 
in our structural relaxation.
Among various $n$ values, we present results for $n$ = 6.7 (Eu$^{2.3+}$) 
as an optimized bulk structure at room temperature, 
and  $n$ = 6.2 (Eu$^{2.8+}$) as an optimized structure at low temperature (below 30{ }K), 
following the suggested mean valencies 
from the experimental reports in Ref.~\cite{sampath,mimura}. 
The 12$\times$12$\times$12 $k$-mesh was adopted for atomic position relaxations, 
while a dense $k$-mesh of 21$\times$21$\times$21 was used for accurate total energy calculations 
to determine the energetically stable structure. 
All relaxed lattice parameters obtained at each $n$ value in \eups~are shown in Fig.~\ref{str}(c), 
which agree well with the experimentally observed lattice parameters 
in dependence of the Eu valence states~\cite{palenzona,mitsuda2,kliemt2022}.
Specifically, we mention the fully relaxed lattice parameters 
are $a$ = 4.214{ }\AA\ and $c$ = 9.895{ }\AA\ at $n$ = 6.7 (Eu$^{2.3+}$), 
and $a$ = 4.128{ }\AA\ and $c$ = 9.886{ }\AA\ at $n$ = 6.2 (Eu$^{2.8+}$).
For surface-sensitive electronic structure calculations,
we constructed Eu-terminated 1$\times$1$\times$4 slab structures of \eups~with a vacuum layer 
of 15{ }\AA~using our optimized bulk structure at room temperature 
and then relaxed the atomic positions of the four layers close to the surface
using the {\it open-core} approximation with $n$ = 6.7 for Eu 4$f$ within LDA in FPLO.

All electronic and magnetic properties were calculated with WIEN2k 
including spin-orbit coupling (SOC) and correlation effects ($U$, $J_H$) 
so as to deal with the localized nature of Eu 4$f$ orbitals. 
We fixed $U$ = 6{ }eV and the Hund's coupling $J_H$ = 1{ }eV for the Eu atom.
The $k$-mesh sampling was 17$\times$17$\times$17 for the bulk states,
and 17$\times$17$\times$1 for the slab structure.
The size of the basis set was determined by the value of $R_{mt}K_{max}$ = 9.0 
with the muffin-tin radius of 2.5(Eu), 2.4(Pd), 1.95(Si), and 2.25(Ge) 
in atomic units.

To describe the divalent states (Eu$^{2+}$) in DFT, the inclusion of spin degrees of freedom 
in spin-polarized calculations was taken into account for simulation purposes, 
even though \eups~at room temperature does not magnetically order.
We assumed ferromagnetic (FM) spin order with an easy axis parallel to the $z$-axis.
In addition, for the comparison of the electronic structures between \eups~and \eupg,
an A-type AFM spin order where the Eu magnetic ions
are ferromagnetically aligned within the $ab$ plane and antiferromagnetically aligned 
between consecutive planes was set. 
The corresponding magnetic space group is $P_I4/mnc$ (No.128.410).
For these calculations, we also fully relaxed the \eupg~crystal bulk structures as we did in \eups.
The fully relaxed lattice parameters in \eupg~at $n$ = 7 are
$a$ = 4.344{ }\AA\ and $c$ = 10.217{ }\AA, which are also in good agreement with the experiment
($a_{exp}$ = 4.3764{ }\AA\ and $c_{exp}$ = 10.072{ }\AA)~\cite{onuki2020}.

We would like to note that within DFT it is difficult to trace the valence transition 
in one single calculation due to the required different treatment of non-magnetic 4$f^6$ Eu 
and magnetic 4$f^7$ Eu. 
The way we approach the valence transition in what follows is therefore by investigating 
first the two limiting valency situations that require within DFT a different type of calculation, 
and, we then approach the transition from both sides as a function of volume variation, 
analyzing the changes in the electronic and magnetic properties of \eups.

Single crystals of \eups~were grown using the Czochralski method 
according to the procedure described in Ref.~\cite{kliemt2022}.
Angular-resolved photoemission measurements (ARPES) on the (001) surface were performed 
at the $1^3$ ARPES instrument at BESSY~II~\cite{onecube}. 
To prepare a clean surface the samples were cleaved $in situ$ under ultra-high vacuum conditions 
at a temperature of 41{ }K. 
In the experiment the energy resolution is better than 50{ }meV, 
the angular resolution is better than 0.2{ }deg.

\section{Structural Details}

\eups~crystallizes in a tetragonal body-centered ThCr$_{2}$Si$_{2}$-type~\cite{thcr2si2} 
structure with space group $I4/mmm$ (No.139)~\cite{palenzona}.  
It consists of layers of edge-sharing PdSi$_4$ tetrahedra intercalated 
between Eu planes, as shown in Fig. \ref{str}(a). 
Eu, Pd and Si are at Wyckoff positions 2$a$, 4$d$, and 4$e$ 
($z_\mathrm{rel}$ = 0.3779 for Eu$^{2.3+}$ and 0.3818 for Eu$^{2.8+}$) respectively).
Similar to Si, Ge in \eupg~also sits on the 4$e$ site with $z_\mathrm{rel}$ = 0.3715 
for Eu$^{2+}$.

\begin{figure}[t]
\vskip 2mm
\includegraphics[width=1.0\columnwidth]{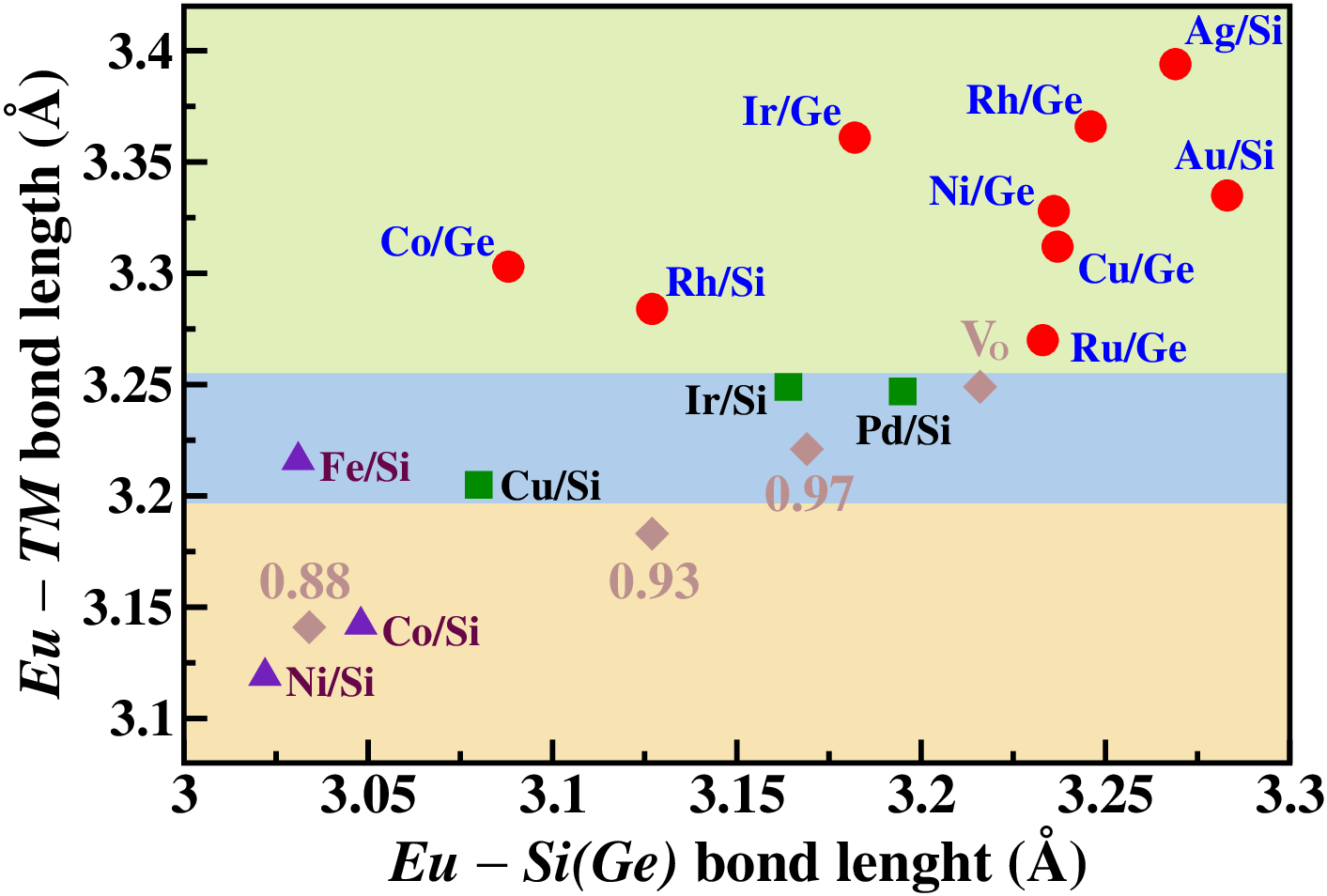}
\caption{ Classification of tetragonal Eu compounds Eu$TM_{2}X_{2}$
according to their experimentally reported Eu-$TM$ and Eu-$X$ bond lengths, where 
$TM$ = Fe, Co, Ni, Cu, Ru, Rh, Pd, Ag, Ir, Au, and $X$ = Si, Ge.
Data were obtained from, respectively, $TM$/$X$ = Ni/Si~\cite{NiSi}, Co/Si~\cite{CoSi},
Fe/Si~\cite{FeSi_NiGe}, Cu/Si~\cite{CuSi}, Co/Ge~\cite{CoGe}, Ir/Si~\cite{IrSi}, 
Rh/Si~\cite{RhSi}, Pd/Si~\cite{PdSi}, Ir/Ge~\cite{RuGe_IrGe,IrGe2}, Ru/Ge~\cite{RuGe_IrGe,RhGe-RuGe}, 
Cu/Ge~\cite{CuGe,CuGe2}, Ni/Ge~\cite{FeSi_NiGe,CoGe}, Rh/Ge~\cite{RhGe,RhGe-RuGe}, Au/Si~\cite{AuSi_AgSi,AuSi2}, 
Ag/Si~\cite{AuSi_AgSi,AgSi2}.
Systems with divalent Eu$^{2+\delta}$ states at low temperatures are shown by red circles, 
whereas purple triangles denote trivalent Eu$^{3-\delta}$ compounds. 
Green squares indicate compounds which undergo a valence transition by varying temperature.
Brown rhombuses indicate bond lengths of relaxed \eups~at a given volume 
with respect to the relaxed one ($V_o$) calculated for $n$ = 6.7 (see Sec. III).
Below 0.93$V_o$, bond lengths become similar to those of trivalent Eu compounds. 
}
\label{bondlength}
\end{figure}

The valence transition in \eups~is accompanied by a volume contraction
where the lattice parameter $a$ shrinks from $a$ = 4.24{ }\AA~at room temperature
to $a$ = 4.16{ }\AA~below 50{ }K. 
To study the relation between volume contraction 
(and corresponding atomic bond-lengths shortening) 
with the valence transition, we collected in Fig.~\ref{bondlength} crystal information 
of ThCr$_2$Si$_2$-type tetragonal Eu-based compounds Eu$TM_{2}X_{2}$ 
whose valence states are confirmed experimentally.
Figure~\ref{bondlength} illustrates two kinds of bond lengths between Eu 
and the transition metal ($TM$) ion, and between Eu 
and the carbon-group ($X$) ion, respectively.
Red circles indicate tetragonal Eu compounds that show magnetic ground states 
(e.g., with divalent Eu$^{2+\delta}$), whereas purple triangles are nonmagnetic 
(Eu$^{3-\delta}$) compounds.  
Systems undergoing a valence transition when temperature decreases are marked 
by green squares. 
The following compounds undergo a pressure-induced valence transition 
at pressures of about 1{ }GPa for EuRh$_2$Si$_2$~\cite{mitsuda3}, 
2{ }GPa for EuNi$_2$Ge$_2$~\cite{nakamura}, and 3{ }GPa for 
EuCo$_2$Ge$_2$~\cite{dionicio}. 
Furthermore, recent surface-sensitive photoemission experiments 
on tetragonal EuIr$_{2}$Si$_{2}$, marked by a green square in Fig.~\ref{bondlength}, 
reveal divalent Eu$^{2+}$ states in the surface region with two-dimensional 
ferromagnetic order at low temperatures, while in the bulk Eu is almost trivalent 
and has a nonmagnetic ground state.~\cite{schulz,usachov2020}.

Interestingly, EuCu$_2$Si$_2$, denoted by a green square in Fig.~\ref{bondlength}, 
has been reported to show a valence transition by lowering temperature~\cite{Bauminger,Patil},
however, while the authors of Ref.~\cite{pagliuso} reported 
the appearance of Eu antiferromagnetism at 10{ }K in single crystals, 
the authors of Ref.~\cite{kawasaki} suggested that the appearance of volume contractions 
and corresponding change of Eu valence states at low temperatures originated 
from the crystallization method and crystal defects. 
These authors confirmed the presence of trivalent states 
on a single crystal EuCu$_2$Si$_2$ at low temperatures.
Actually, the single crystal from Ref.~\cite{pagliuso} has a larger volume by 
about 3{ }\% compared to the samples showing trivalent states~\cite{kawasaki}. 
Recently, it was found experimentally that bond lengths 
and valence transition temperature in \eups~can also change 
depending on the amount of disorder in the Pd-Si layer~\cite{kliemt2022}.

Summarizing the above observations, the Eu valence state 
in these tetragonal Eu-based compounds is intimately related to the value 
of Eu-$TM$ and Eu-$X$ bond lengths in the systems. 
In other words, unlike the previous usual way to refer to systems' volume,
the relation between Eu valency and the bond lengths is an interesting parametrization 
and should be taken into account when it comes to the valence state of Eu 
on tetragonal Eu-based compounds.
In what follows we concentrate on \eups~and analyze via {\it ab initio} DFT
the Eu valence transition under volume reduction.
It should be noted that in a unit cell volume, there are various atoms involved, 
and therefore bond lengths and angles may change differently with a given volume change.

\section{Electronic Structure}

\begin{figure*}[t]
\includegraphics[width=2.0\columnwidth]{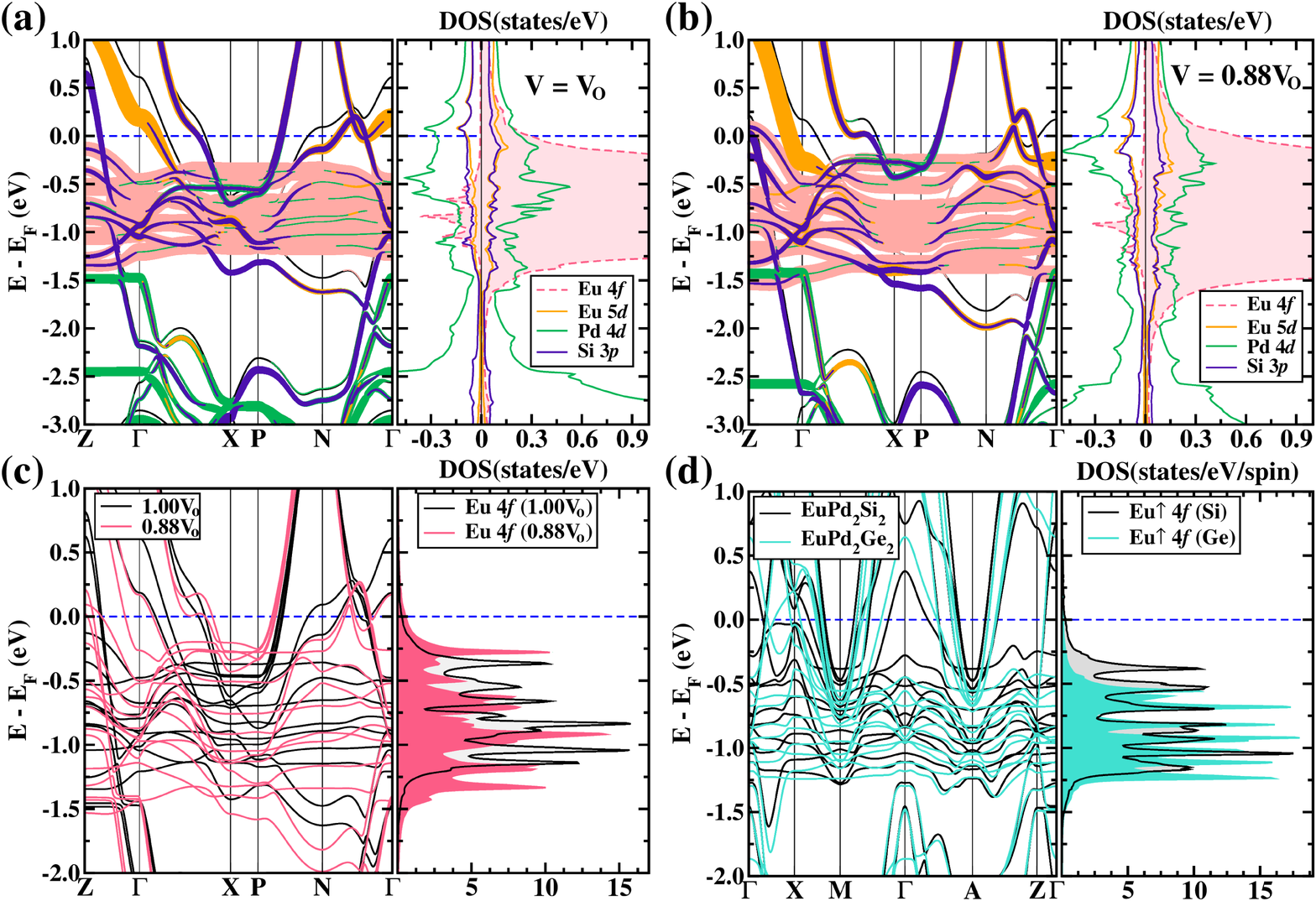}
\caption{
Fat band electronic structures and orbital projected DOSs 
obtained from LSDA+SOC+$U$ for (a)-(c) \eups~and (d) \eupg.
(a) FM band structures for \eups~at $V_o$ ($n = 6.7$ relaxed structure, see Sec. III)
and (b) at a volume of 0.88$V_o$. 
At $V_o$ in (a), the occupied spin-up Eu 4$f$ bands in the FM results
are located just below the Fermi level.
The energy position of the Eu 4$f$ states is in good agreement 
with photoemission results, as shown in Fig. \ref{surf}(b).
At 0.88$V_o$, the $c$-$f$ hybridization is enhanced leading 
to a larger bandwidth of 4$f$ bands in (b).
The weight of Eu 4$f$ band characters (a-b) was scaled by a factor of 0.25 for clarity. 
(c) Overlapped band structures and Eu 4$f$ orbital projected DOSs 
within LSDA + SOC + $U$ for \eups~at $V_o$ (black color) and at 0.88$V_o$ (light red color). 
The volume reduction in \eups~with shorter bond lengths induces 
an enhanced $c$-$f$ hybridization leading to more dispersive bands 
and a larger bandwidth of Eu 4$f$.
(d) Overlapped band structures and orbital projected DOSs for \eups~(black color) 
and \eupg~(cyan color) at $V_o$ both calculated in the A-AFM state. 
Longer bond lengths in \eupg~lead to a narrower band width reflecting 
more localized Eu 4$f$ states compared to \eups.
}
\label{bulkband}
\end{figure*}

\subsection{Calculations for bulk \eups~}
We start by examining the description of Eu 4$f$ states in the context of competing 
Coulomb interaction, crystal field environment, and spin-orbit coupling at ambient pressure. 
As it is known from atomic physics~\cite{woodgate1970}, 
the description of the electronic structure of an atom depends 
on the hierarchy of the involved interactions.
In the case that the Coulomb interaction between the electrons is stronger than 
the spin-orbit interactions in each of them, then the total angular momentum $L = \sum_i l_i$ 
and total spin $S = \sum_i s_i$ of the electron system, where $l_i$ and $s_i$ are, respectively, 
the angular momentum and spin of each individual electron, couple to a total $J$. 
This situation corresponds to the $LS$ description. 
If, however, the individual electron coupling via the spin-orbit interaction is 
stronger than the Coulomb interaction between electrons $U_{ee}$, 
then the individual total momenta $j_i = l_i+s_i$, couple to a total $J = \sum_i j_i$ 
which corresponds to a $jj$ description. 
Usually the previous description is valid for light atoms, 
while the second one is more appropriate for heavy atoms. 
In 4$f$ systems $U_{ee}$ is usually larger than the SOC constant $\xi_{SOC}$ 
and the $LS$ coupling scheme may be more appropriate~\cite{hotta}.
Actually, in a $LS$ description Eu$^{2+}$ (4$f^7$) has  $S = 7/2$, $L = 0$ 
and a total $J = 7/2$. 
Alternatively, in the $jj$ description $j_i$ with $i = 1,2,...,7$ 
can take values $5/2$ and $7/2$ and the total $J$ in the ground state is 
then $J = 7/2$ as well. 
Analogously  Eu$^{3+}$ (4$f^6$) in a $LS$ description has $S = L = 3$ and $J = 0$. 
Considering the $jj$ description, $j_i$, $i = 1,2,...,6$ it results in $J = 0$. 
The differences between the two schemes is perceived when investigating 
the magnetic moments~\cite{hotta} $M = g_J\mu_B J$ where the Land\'{e} factor $g_J$ 
corresponds to the electron gyromagnetic factor $g_J = 2$ in the $LS$ scheme, 
while in the $jj$ coupling $g_J = 8/7$ for Eu$^{2+}$. 
These nuances in the description are important 
when comparing the calculated magnetic properties to the experiment.

Now we analyze the limiting case of the relaxed structure obtained 
for $n = 6.7$ (see Sec. III) corresponding to \eups~at ambient pressure 
and ambient temperature~\cite{mimura}.
Figure \ref{bulkband}(a) shows the LSDA + SOC + $U$ calculated electronic structure and 
corresponding orbital projected densities of states (DOSs) for \eups~where we assumed 
a FM configuration for Eu. 
Due to the half-filled shell (4$f^7$), the fully occupied 4$f$ bands 
in the majority spin channel [colored pink in Fig. \ref{bulkband}(a)] are centered 
around $-$1.0{ }eV below the Fermi level with a bandwidth of about 1{ }eV.
Note that the energy of the calculated $4f$ states corresponds to the energy position 
of the Eu$^{2+}$ final-state multiplet seen in photoemission, Fig.~\ref{surf}(b).

Furthermore, in this energy range, itinerant conduction states of Pd 4$d$, Si 3$p$, and
Eu 5$d$ are present, which hybridize with the half-filled Eu 4$f$ bands.
Specifically, there is a hole pocket with dominant Eu 5$d$ 
character at the $\Gamma$ point, as well as electron pockets with
dominant Pd 4$d$ and Si 3$p$ characters at around the $X$ and $P$ points.
The calculated total spin moment ($M_S$) is 6.95{ }$\mu_B$/Eu 
with a negligible angular moment of $-$0.024{ }$\mu_B$/Eu, 
which is in good agreement with the values of $S$ = $\frac{7}{2}$ 
and $L$ = 0 expected from the Hund's rule in the divalent Eu$^{2+}$ state in the $LS$ scheme.
Specifically, the occupation numbers of each orbital are 6.716 for Eu 4$f$, 8.016 for Pd 4$d$, 
0.815 for Si 3$p$, and 0.331 for Eu 5$d$ orbitals.
In contrast to \eups, EuPd$_2$Ge$_2$ does not undergo a valence transition 
at low temperatures but an AFM transition at $T_\mathrm{N} = 17${ }K~\cite{onuki2020}. 
This is directly related to the fact that Eu-Pd / Eu-Ge bond lengths 
in EuPd$_2$Ge$_2$ are longer than Eu-Pd / Eu-Si bond lengths 
in \eups~and, following Fig.~\ref{bondlength}, Eu$^{2+}$ states are expected.
Our relaxed bond lengths of Eu-Pd / Eu-Ge in EuPd$_2$Ge$_2$ differ by 3.2 / 3.9{ }\% from 
those of Eu-Pd / Eu-Si in \eups.
Figure \ref{bulkband}(d) illustrates band structures and Eu 4$f$ orbital projected DOS
of A-AFM \eups~(black) and  EuPd$_2$Ge$_2$ (cyan) within LSDA + SOC + $U$. 
Due to the longer bond lengths, 4$f$ states in EuPd$_2$Ge$_2$ near the Fermi level 
are less dispersive and have a smaller bandwidth than those of \eups.

We consider now the relaxed \eups~structure at $n$ = 6.2 (see Sec. III). 
To obtain the electronic structure we performed calculations 
within LSDA + SOC + $U$ ($U = 6${ }eV, $J_\mathrm{H} = 1${ }eV), 
with zero initial magnetization for the Eu atom.
Both combined effects, SOC and correlation ($U$) shift down the occupied sixfold 
$| j$ = $\frac{5}{2} \rangle$ states to about 6.4{ }eV below the Fermi level 
and show almost no dispersion, whereas the empty eightfold $| j$ = $\frac{7}{2} \rangle$ 
states are located around 1.6{ }eV above the Fermi level, not shown here.
With the choice of $U$ and $J_\mathrm{H}$ values above, 
the energy position of the occupied $| j$ = $\frac{5}{2} \rangle$ states 
coincides with the position of the trivalent Eu states 
in photoemission measurements [compare to Fig. \ref{surf}(b)]. 
Note, however, that in Fig. \ref{surf}(b), due to the multiplet nature of trivalent Eu, 
two separated main peaks are observed at about 7{ }eV below the Fermi level. 
This feature is not captured in the DFT calculations, 
due to the limitations of the method to describe manybody multiplet states.

\subsection{Role of volume reduction on the hybridization}

We analyze now the effect of volume reduction on the valence transition 
from divalent Eu to trivalent Eu in \eups.
As mentioned above, this system undergoes a valence transition
by lowering temperature or by increasing pressure~\cite{sampath,adams}.
Both effects reduce the volume in \eups, implying a shortening of the Eu-Pd 
and Eu-Si bond lengths. 

To investigate how the volume reduction affects the electronic structure 
and the valence state of \eups, we obtained a few bulk \eups~structures 
with a smaller volume than our fully relaxed one ($V_o$) at $n = 6.7$
by relaxing them using the {\it open-core} approximation as implemented in FPLO within LDA.
Our fully relaxed lattice parameters at $n$ = 6.7 (Eu$^{2.3+}$), 
which correspond to the room temperature bulk structure at ambient pressure
are $a$ = 4.214 and $c$ = 9.895{ }\AA~as given in the Sec. III 
with bond lengths of Eu-Pd = 3.25{ }\AA~and Eu-Si = 3.22{ }\AA. 
Note that the Eu-Si bond length is in good agreement with our value determined 
in Ref.~\cite{kliemt2022}.
These bond lengths are marked by a brown rhombus with $V_o$ in Fig. \ref{bondlength}.
These values are slightly larger than the reported ones in the literature~\cite{PdSi}. 
Also plotted in Fig. \ref{bondlength} are the results
for relaxed structures at a volume of 0.97$V_o$, 0.93$V_o$, and 0.88$V_o$.
We note that the lattice parameter $a$ for the crystal structure relaxed at 0.97$V_o$ 
is similar to the experimentally reported one at a temperature below 30{ }K and 0{ }GPa 
when the system has undergone the valence transition and has experienced a volume contraction.
The values of bond lengths are in the same range as those for compounds 
that undergo a valence transition (marked by green squares in Fig. \ref{bondlength}). 

\begin{figure}[t]
\vskip 2mm
\includegraphics[width=1.0\columnwidth]{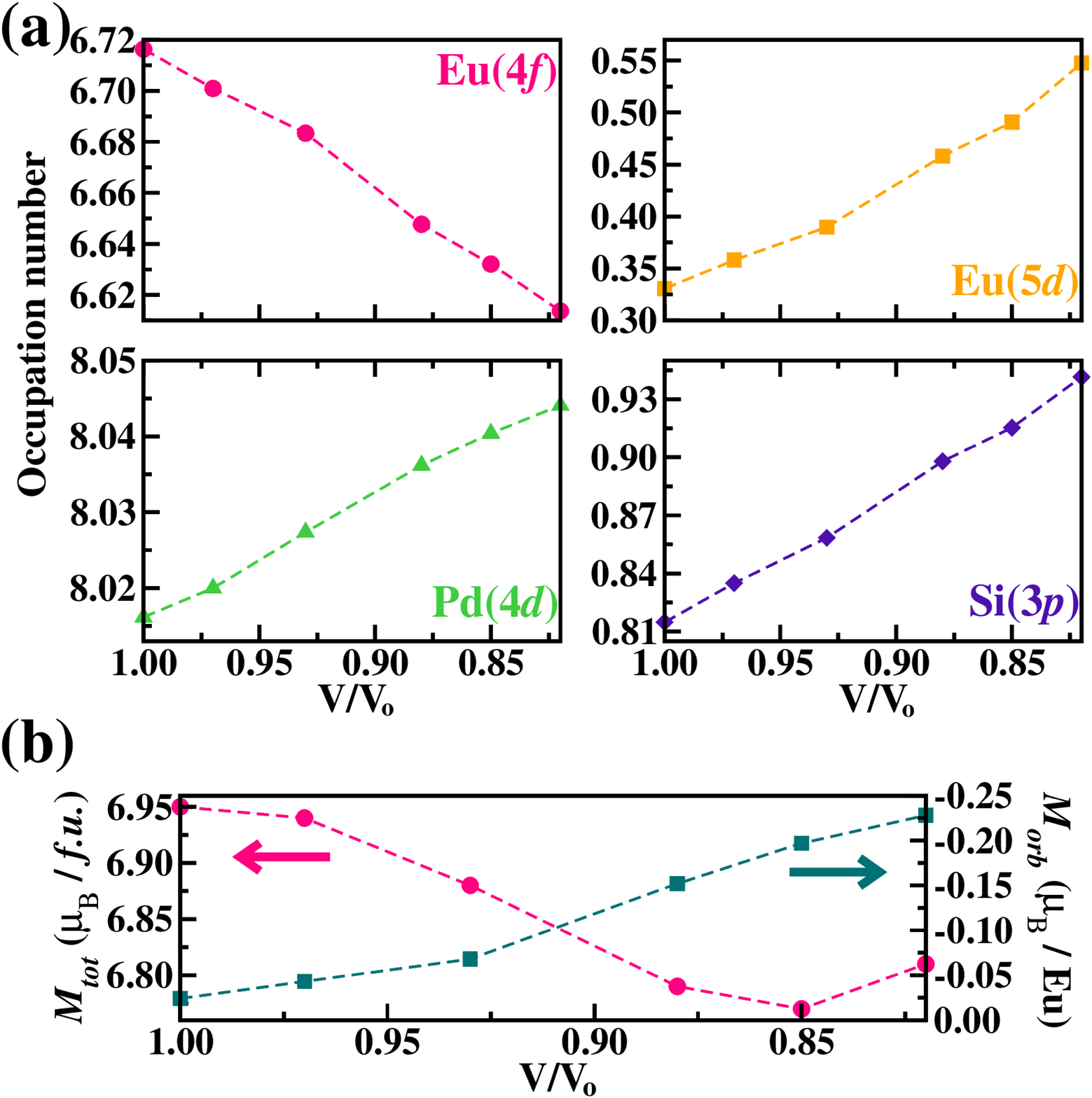}
\caption{
(a) Occupation number of each orbital as a function of volume reduction 
of \eups~within LSDA + SOC + $U$ where $V_o$ corresponds to the structure relaxed  
with $n = 6.7$ (see Sec. III).
When the volume is reduced, the number of occupied Eu 4$f$ states decreases, 
whereas there is an increase in the other states.
(b) Corresponding total spin moment and orbital moment of Eu as a function of volume compression.
A decrease of volume results in a decrease of the total spin moment, 
whereas the orbital moment of Eu increases in magnitude.
}
\label{occ}
\end{figure}

Figure \ref{bulkband}(a) and \ref{bulkband}(b) show the fat band electronic structure 
and orbital projected DOSs of FM \eups~in LSDA+SOC+$U$ 
at a volume of $V_o$ and 0.88$V_o$, respectively. 
The majority 4$f$ bands are located just below the Fermi level.
Compared to the $V_o$ results, at 0.88$V_o$ the 4$f$ bandwidth is somewhat larger 
and more dispersive with an enhanced $c$-$f$ hybridization. 
This can be better observed in Fig. \ref{bulkband}(c) where
we superimposed both electronic structure contributions.
Specifically, a reduction of volume leads to a decrease of the occupation number of Eu 4$f$ states,
whereas the occupation number in the other orbitals goes up, as shown in Fig. \ref{occ}
where we display the occupation of Eu 4$f$, Pd 4$d$, Si 3$p$, and Eu 5$d$ states 
as a function of volume.
This is directly related to a shortening of the Eu-$TM$ and Eu-$X$ bond lengths,
(see brown rhombuses in Fig.~\ref{bondlength})
and an enhanced hybridization between Eu 4$f$ and the itinerant conduction states 
(Eu 5$d$, Pd 4$d$, and Si 3$p$).
Further, under volume reduction the total spin moment decreases, 
whereas the orbital moment increases in magnitude [see Fig. \ref{occ}(b)].
At 0.88$V_o$, the orbital moment of Eu becomes $-$0.152{ }$\mu_B$/Eu 
which is a consequence of a slight decrease of the occupation of a mostly $|3,3\rangle$ state,
whereas the occupation of the other states remains nearly unchanged.
Furthermore, the orbital moment increases to $-$0.228{ }$\mu_B$/Eu at 0.82$V_o$.
This behavior shows a tendency to follow the Hund's rules for Eu$^{3+}$ 
which is $L$ = $S$ = 3 and $J = |S - L|$ = 0.

We conclude this section by the observation that an $LS$ scheme is more suitable to describe the
electronic structure of \eups~with Eu in the nearly divalent state,
while in the case of Eu in a nearly trivalent state corresponding to the reduced volume case, 
the $jj$ coupling scheme seems more appropriate, as discussed in Ref.~\cite{hotta}. 
The reason for that is the relative changes in the ratio $\xi_{SOC}$/$U_{ee}$~\cite{hotta}  
where $\xi_{SOC}$ is the spin-orbit coupling strength and $U_{ee}$ the Coulomb interaction.  
This ratio increases at reduced volumes, as happens while lowering temperature,
due to an increased $c$-$f$ hybridization.

\subsection{Photoemission from the $4f$ shell}
Recently, some of us~\cite{schulz} reported two-dimensional ferromagnetism 
at a temperature below 48{ }K in a single Eu layer located below the iridium-silicide surface 
of tetragonal EuIr$_2$Si$_2$, which shows a temperature-driven valence crossover 
(marked by a green square in Fig. \ref{bondlength}), 
whereas bulk regions display no magnetism due to the presence of Eu$^{3+}$.
In a later experiment, surface ferromagnetism was observed at the Eu-terminated surface of EuIr$_2$Si$_2$ as well~\cite{usachov2020}. 
Figure \ref{surf}(b) shows an ARPES spectrum acquired from the (001) surface 
of a freshly cleaved EuPd$_2$Si$_2$ single crystal at a temperature of 41{ }K. 
Although the compound was first synthesized decades ago, only recently large single crystals 
are available~\cite{kliemt2022}, enabling ARPES measurements. 
On the right-hand side of Fig. \ref{surf}(b) the corresponding angle-integrated spectrum is plotted. 
To maximally enhance the emission from the $4f$ shell over contributions from the valence band, 
we used a photon energy of 145{ }eV which corresponds to the maximum of the $4d\rightarrow 4f$ 
Fano-Beutler resonance of Eu$^{3+}$. 
Note that at the given photon energy the $4f$ emission of Eu$^{2+}$ is resonantly enhanced as well. 
In the spectrum, three dominating non-dispersive $4f$ features are present, 
which are well established 
for mixed-valent Eu systems \cite{kawasaki, schulz, usachov2020, kawasaki2021} 
and were reported for photoemission experiments on \eups{}, also \cite{martensson}. 
Those are (1) the straight line at the Fermi level represents 
the $4f^7\rightarrow 4f^6$ final-state multiplet of Eu$^{2+}$ in bulk-like layers; 
(2) the most intense line at a slightly higher binding energy of about 
1{ }eV is the surface-core-level shifted $4f$ emission of Eu$^{2+}$ at the surface; 
(3) the broad structure consisting of several lines between 6 and 10{ }eV forms 
the $4f^6\rightarrow 4f^5$ final-state multiplet of Eu$^{3+}$ in bulk-like layers. 
The simultaneous observation of both the Eu$^{2+}$ and Eu$^{3+}$ final-state multiplets reflects 
the mixed-valent properties of Eu in this compound.

\subsection{Slab calculations} 
 
\begin{figure}[t]
\vskip 2mm
\includegraphics[width=\columnwidth]{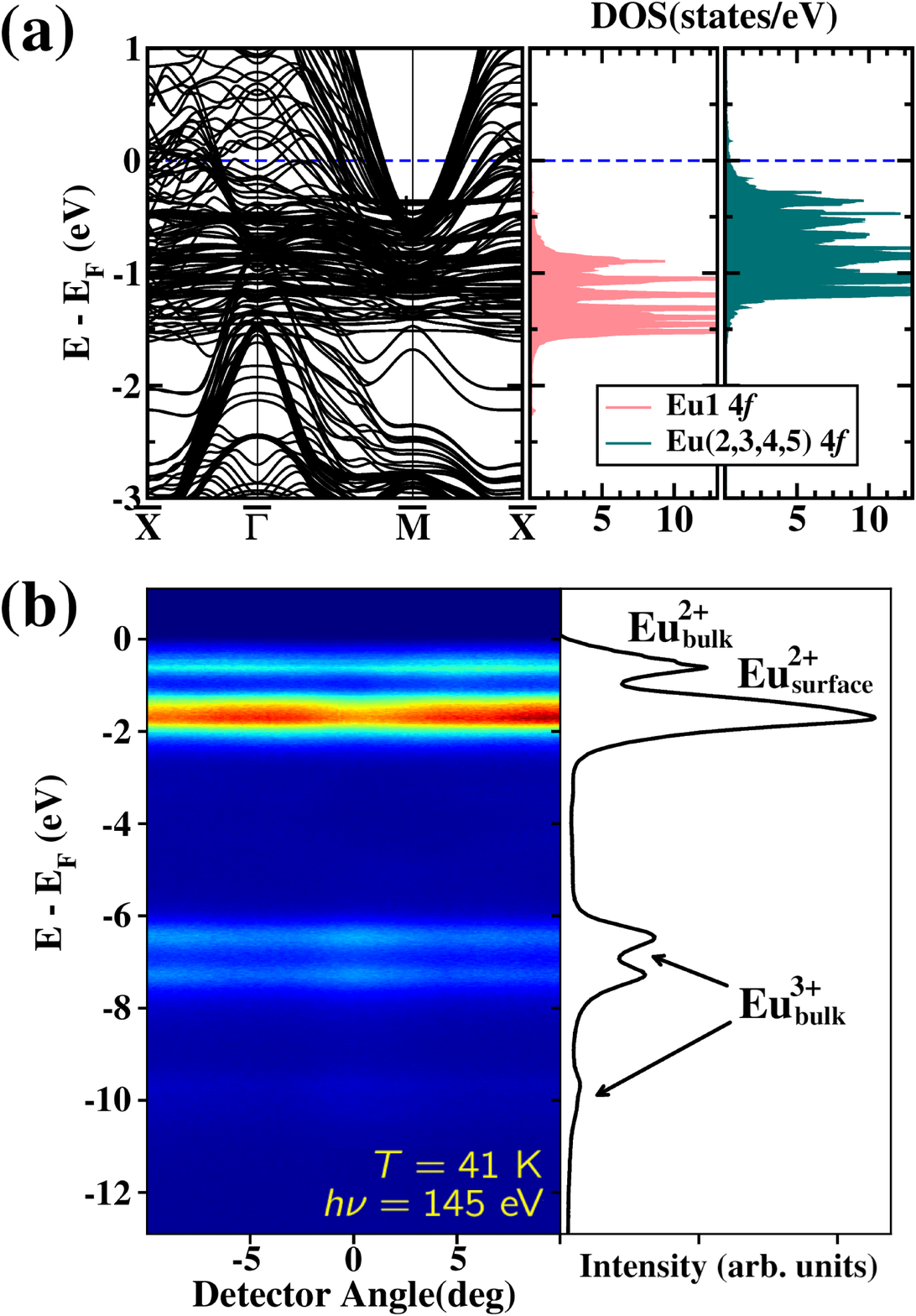}
\caption{(a) LSDA + SOC + $U$ electronic band structure 
and 4$f$ orbital projected DOS of the Eu-terminated 1$\times$1$\times$4 slab structure 
with a spin configuration where all Eu atoms were set to be of FM order.
(b) ARPES spectrum acquired with $h\nu = 145${ }eV at a temperature of 41{ }K 
from the (001) surface of \eups~for a mixture of Si- and Eu-terminated areas, 
that shows the Eu $4f$ emission. 
On the right, the angle-integrated PES spectrum is given. 
The DFT calculations capture the distinct surface versus bulk divalent Eu states as observed 
in the PES experiments in the energy region [$-$3{ }eV,0].
}
\label{surf}
\end{figure}

To compare our DFT calculations to the surface-sensitive ARPES measurements presented 
in the previous section, we constructed a 1$\times$1$\times$4 
Eu-terminated slab geometry as described in the Sec. III. 
Such a geometry allows to disentangle the surface from the bulk states.
Due to the symmetric geometry, where the space group of this slab structure is $P4/mmm$, 
there are five inequivalent Eu atoms in the unit cell of our slab. 
Numbering these Eu atoms in relation to proximity to the surface, 
they are Eu1 through Eu5 with Eu1 at the surface and Eu5 situated  
furthest from the surface, i.e., located at the center of the unit cell. 
For the calculation of the electronic structure, 
we considered FM configurations of these Eu atoms as in our bulk calculations
and concentrate therefore on the description of divalent Eu.
Figure \ref{surf}(a) illustrates the (001)-projected band structure and 
Eu 4$f$ orbital resolved DOS of the slab structure described 
above for the FM spin configuration in LSDA + SOC + $U$.
All nearly divalent Eu 4$f$ states appear at around 1{ }eV below the Fermi level. 
We observe that 4$f$ states of Eu2, Eu3, Eu4, Eu5 states belonging to the bulk are 
much closer to the Fermi level than the surface Eu1 states, in agreement with our PES results 
in the energy range [$-$3{ }eV,0], shown in Fig. \ref{surf}(b), 
where a peak from Eu$^{2+}$(surface) is farther from the Fermi level 
than the weight corresponding to bulk Eu.

\section{Conclusions}

In this work we investigated from first principles the microscopic mechanism 
of the valence transition in the test-bed system \eups~which is known 
to undergo a valence transition from nearly divalent Eu to nearly trivalent Eu 
upon lowering the temperature. 
By making use of the observation that the valence transition is accompanied 
by a volume contraction, we studied the evolution and occurrence of the valence transition 
by a combination of (i) density functional theory calculations 
where we considered volume contracted structures, and (ii) photoemission measurements taken 
at T = 41{ }K with $h\nu = 145${ }eV that were used to benchmark our calculations.
Our analysis of the electronic and magnetic properties of \eups\ when approaching 
the valence transition showed an enhanced $c$-$f$ hybridization between localized Eu 4$f$ states 
and itinerant conduction states (Eu 5$d$, Pd 4$d$, and Si 3$p$) 
where an electronic charge redistribution, that we quantified, takes place.
The change in the electronic structure was shown to be intimately related to 
the volume reduction where Eu-Pd(Si) bond lengths shorten. 
As the bond lengths get shorter the occupation number of the Eu $4f$ states decreases 
whereas that of the conduction states increases.  
For the transition to happen, we observe a delicate balance between electronic bandwidth, 
crystal field splitting, Coulomb repulsion, Hund's coupling, 
and spin-orbit coupling that we trace with our calculations.

Further, our DFT \eups~bulk and Eu-terminated slab results are in good agreement 
with our surface sensitive photoemission experiments reproducing 
the presence of divalent Eu states near the Fermi level coming from surface/bulk Eu 
and of trivalent Eu states at high binding energies developing from bulk Eu.

With this study we also explored the limits of density functional theory 
and the choice of exchange correlation functionals to describe such a phenomenon 
as valence transition.

\section{Acknowledgments}
We thank Daniel Khomskii, Michael Lang, Igor Mazin and Bernd Wolf for useful discussions.
Y.-J. S., K.K., C.K., and R.V. acknowledge support by the Deutsche Forschungsgemeinschaft 
(DFG, German Research Foundation) for funding through TRR 288 -- 422213477 (projects A03, A05). 
S.S. acknowledges DFG support through Grant No. KR3831/5-1. 
We thank the Helmholtz-Zentrum Berlin f\"ur Materialien und Energie 
for the allocation of synchrotron radiation beamtime.

\newpage

\bibliography{ref,ref_DFT}

\end{document}